\documentclass[]{KapProc}

\usepackage{psfig}\normallatexbib

\begin{document}
\articletitle{Optical qubit using linear elements}
\author{Matteo G. A. Paris ({\tt paris@unipv.it})}
\affil{Quantum Optics Group, Unit\`a INFM and 
Dipartimento di Fisica ``A. Volta'' \\ Universit\`a di Pavia, 
via Bassi 6, I-27100 Pavia, ITALY }
\begin{abstract}
A conditional scheme to prepare optical superposition of the vacuum and
one-photon states using linear elements (beam splitters and phase-shifters) 
and avalanche photodetectors is suggested. 
\end{abstract}
In recent years, the quantum engineering of light have received much attention, 
which is mainly motivated by the potential improvement offered by 
quantum mechanics to the manipulation and the transmission of information. 
In particular, some conditional schemes have been suggested to prepare 
superpositions. Among these we mention the Fock filtering by an active 
Fabry-Perot cavity \cite{fck}, the displacing/photon-adding scheme 
of Ref. \cite{dak}, and the so-called optical state truncation \cite{tru}. 
In this paper we describe a partially interferometric conditional scheme to 
prepare any desired superposition $a_0 |0\rangle + a_1 |1\rangle$ of the 
vacuum and one-photon states using only linear optical components and 
avalanche photodetectors. For a fully interferometric setup and for more 
details we refer the readers to Ref. \cite{opq}. \par
\begin{figure}[ht]
\psfig{file=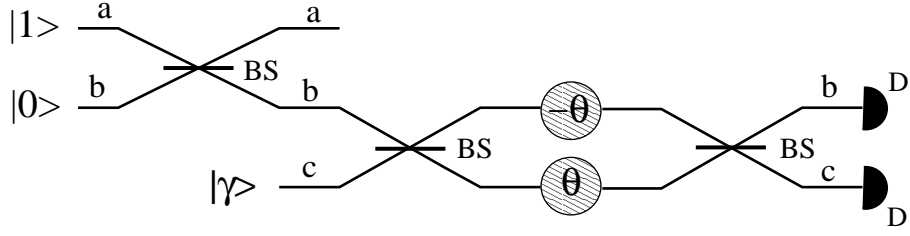,width=12cm}
\caption{Schematic diagram of the setup for the generation of optical 
superpositions. The BS's are balanced beam splitters whereas 
the D's are avalanche photodetectors.\label{f:setup}}
\end{figure}

  
The present scheme (see Fig. \ref{f:setup}) is built by a balanced
beam splitter, fed by one-photon state in the mode $a$, followed by 
a Mach-Zehnder interferometer, with inputs consisting of one of the output 
from the beam splitter (mode $b$) and of an additional mode $c$ excited 
in a weak coherent state. The two modes $b$ and $c$ exiting the 
interferometer are detected, and the situation in which a click is 
observed in one mode, and no clicks are seen in the other one, 
corresponds to the (conditional) preparation of a superposition
of the vacuum and one-photon states in the mode $a$. 
The amplitudes in the superposition may be tuned by varying the 
internal phase-shift of the interferometer and the amplitude of
the coherent state. \par
After the first beam splitter, the joint state of the modes $a$ and $b$ is 
given by the superposition $$|\psi\rangle_{ab} = \frac1{\sqrt{2}}\left[|0
\rangle_a |1\rangle_b + |1\rangle_a |0\rangle_b \right]\:.$$
Mode $b$ then enters the interferometer where it is mixed with mode
$c$ excited in a weak coherent state $|\gamma\rangle$. The evolution operator 
of the interferometer is given by \cite{vis} 
$\hat U (\phi)= \exp \left\{ i \phi \left( b^\dag c + c^\dag b\right)\right\}$, 
where $\phi=\theta/2$. The overall output state is thus given by 
\begin{eqnarray}
|\psi\rangle_{out} = \hat U (\phi) \: |\psi\rangle_{ab} |\gamma\rangle_c
&=& \frac1{\sqrt{2}}\Big[ |1\rangle_a |\gamma\cos\phi \rangle_b 
|\gamma\sin\phi\rangle_c \nonumber  \\  
&+& \sin\phi\: |0\rangle_a\:b^\dag|\gamma\cos\phi\rangle_b 
|\gamma\sin\phi \rangle_c \nonumber \\ &-& \cos\phi \: 
|0\rangle_a |\gamma\cos\phi\rangle_b \; c^\dag|\gamma\sin\phi 
\rangle_c \Big]
\label{out}\;
\end{eqnarray}\par
Let us now analyze the effects of the photodetection of modes $b$ and $c$.
The outcomes from an avalanche detector may be either {\sf YES}, which means
a "click", corresponding to any number of photons, or {\sf NO}, which means
that no photons have been recorded. This kind of measurement is described by
a two-value POM 
\begin{eqnarray}
\hat\Pi_{ N} =  \sum_{p=0}^{\infty} (1-\eta)^p \: |p \rangle\langle p|
\qquad  \hat\Pi_{ Y} = \widehat {\bf I} - \hat\Pi_{ N}
\label{yesno}\;, 
\end{eqnarray}
where $\eta$ is the quantum efficiency, and $\widehat {\bf I}$ denotes the 
identity operator. For high quantum efficiency $\hat\Pi_{N}$ and $\hat\Pi_{Y}$ 
approach the projection operator onto the vacuum state and the orthogonal 
subspace respectively. In this case the event of "no clicks" 
corresponds exactly to the absence of photons. 
In general, the event of observing a click at the detector surveying 
the output mode $b$, and no photons at $c$, 
is  characterized by the probability 
\begin{eqnarray}
P_{YN}[\eta,\gamma,\phi] 
= \hbox{Tr}_{abc}\left[|\Psi_{ out}\rangle\langle
\Psi_{ out}|\; \hat\Pi_{ Y}\:\otimes \hat\Pi_{ N}\right] 
=e^{-\eta|\gamma|^2\sin^2\phi} \times \nonumber \\ 
\Big\{1- e^{-\eta|\gamma|^2\cos^2\phi} + 
\frac{\eta}{2}\Big[e^{-\eta|\gamma|^2\cos^2\phi} 
+ \cos^2\phi \big(\eta |\gamma|^2\sin^2\phi -1\big)
\Big]\Big\}\label{probyn}\;. 
\end{eqnarray}
The corresponding conditional output state is 
\begin{eqnarray}
\hat\varrho_{YN} &=& \frac1{P_{YN}} \hbox{Tr}_{bc}
\left[|\Psi_{out}\rangle\langle\Psi_{out}|\; \hat\Pi_{ Y}\:
\otimes\hat\Pi_{N}\right] = 
\nonumber \\ &=& \frac1{P_{YN}} \Big[ d_{00} \: |0\rangle\langle0| + 
d_{11} \: |1\rangle\langle1| + d_{01} \: |0\rangle\langle1| + d_{01}^* \: 
|1\rangle\langle0| \Big]
\label{rhoyn}
\end{eqnarray}
where the coefficients are given by 
\begin{eqnarray}
d_{11} = \frac12 e^{-\eta|\gamma|^2\sin^2\phi} 
\Big[1-e^{-\eta|\gamma|^2\cos^2\phi}\Big] \quad 
d_{01} = e^{-\eta|\gamma|^2\sin^2\phi} 
\: \frac{\eta \gamma}{2} \: \sin\phi\cos\phi  \nonumber \\ 
d_{00} = \frac12 e^{-\eta|\gamma|^2\sin^2\phi}
\Big[1- (1-\eta)e^{-\eta|\gamma|^2\cos^2\phi}
+\eta\cos^2\phi\big(\eta|\gamma|^2\sin^2\phi-1\big)\Big]\nonumber 
\label{coeffyn}\;.
\end{eqnarray}
The symmetric case of a click observed in the mode $c$ and no clicks
in the mode $b$ leads to an equivalent result, up to the replacement
$\phi\rightarrow \phi + \pi/2$. \par 
Due to non unit quantum efficiency of photodetectors, the conditional output state 
$\hat\varrho_{YN}$ is not a pure state. However, as we will see, there 
are regimes in which $\hat\varrho_{YN}$ approaches the desired superposition. 
In order to compare $\hat\varrho_{YN}$ with the ideal 
conditional output 
$|\psi_{10}\rangle$ we consider the fidelity $F=\langle \psi_{10}|\hat
\varrho_{ yn}|\psi_{10}\rangle$. 
Notice that ideal output corresponds to a conditional photodetection
performed by fully efficient detectors, which are also able to discriminate
among the number of photons.
From previous equations we have
\begin{eqnarray}
F[\eta,\gamma,\phi] &=& \frac1{2P_{YN}}\:\frac{e^{-\eta|\gamma|^2
\sin^2\phi}}{\sin^2\phi+|\gamma|^2\cos^2\phi}
\Bigg\{|\gamma|^2\sin^2\phi \Big(1 - e^{-\eta|\gamma|^2\cos
\phi^2}\Big) \nonumber \\
&&+ 2\eta|\gamma|^2 \sin^2\phi\cos^2\phi+
\sin^2\phi\:\Bigg[1-(1-\eta)e^{-\eta|\gamma|^2\cos^2\phi}
\nonumber \\ &&+
\eta\cos^2\phi\big(\eta|\gamma|^2\sin^2\phi-1\big)\Bigg]\Bigg\}
\label{fid}\;.
\end{eqnarray}
Our aim is to find regimes in which the fidelity of the conditional 
output state is close to unit and, at the same time, the corresponding 
detection probability  $P_{YN}$ does not vanish. \par
\begin{figure}[ht]
\begin{tabular}{cc}
\psfig{file=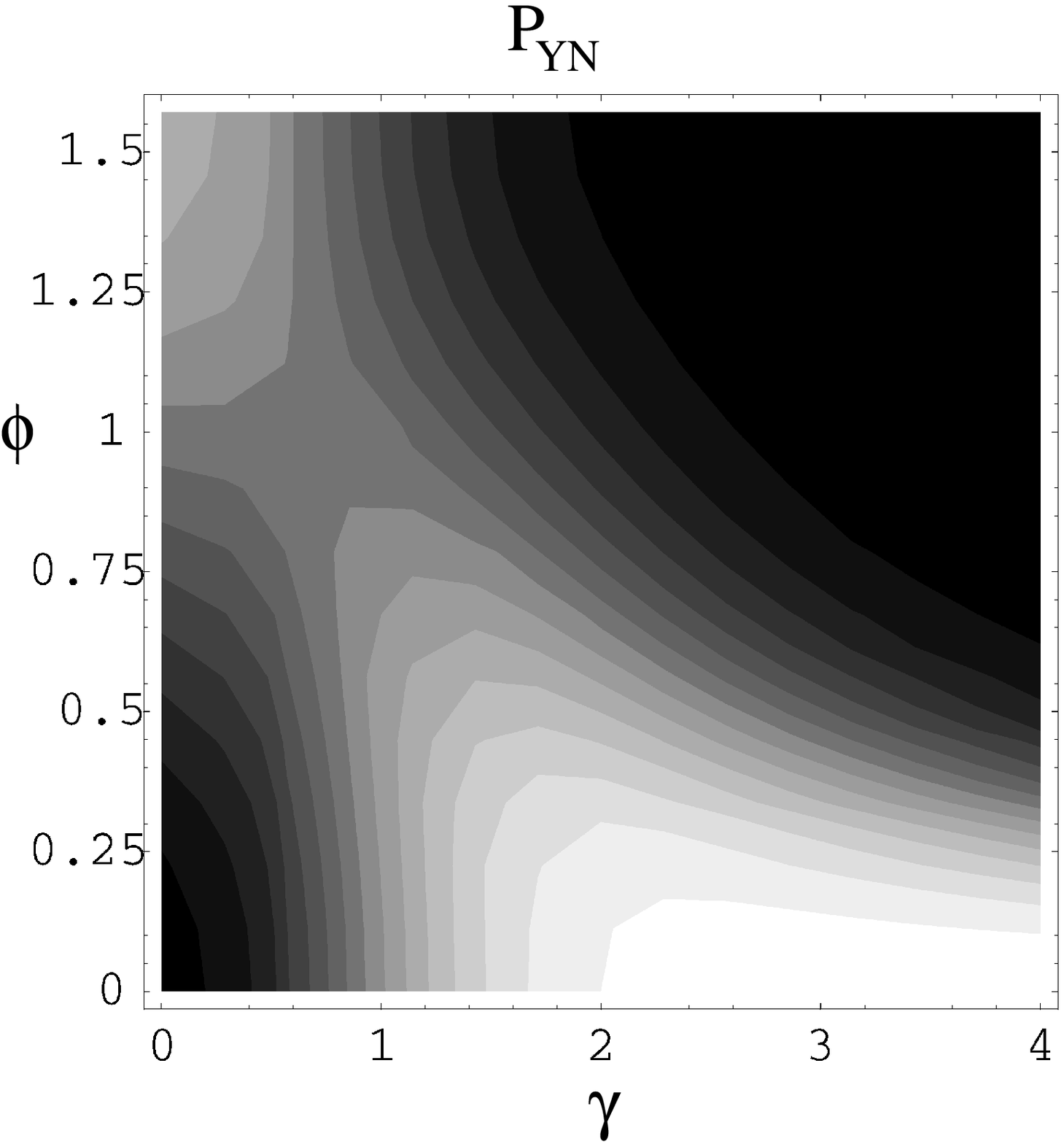,width=45mm}&
\psfig{file=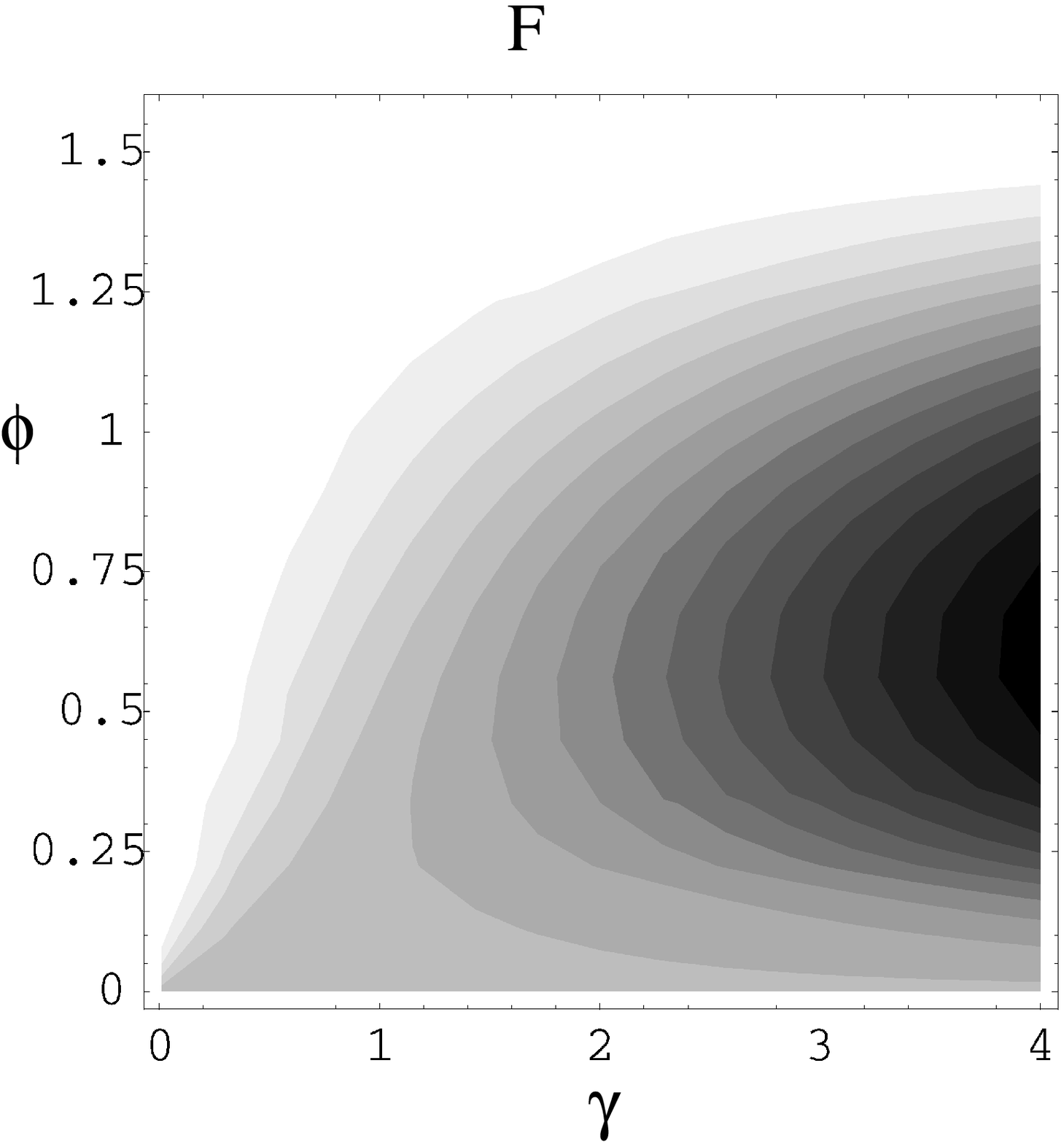,width=45mm}\end{tabular}
\caption{Detection probability and fidelity as a function 
of the coherent amplitude $\gamma$ and the interferometric shift $\phi$ 
for quantum efficiency equal to $\eta=80\%$.\label{f:pf}}
\end{figure}
In Fig. \ref{f:pf} we 
show $P_{YN}[\eta,\gamma,\phi]$ and $F[\eta,\gamma,\phi]$
as a function of $\gamma$ and $\phi$ for quantum efficiency equal to
$\eta=80\%$. As it is apparent from the plot (darker regions correspond 
to lower values of detection probability and fidelity) for a weak coherent 
signal $\gamma \leq 1$ there is, even for non unit quantum
efficiency of the conditional photodetectors, a large range of values of 
$\phi$ corresponding to high fidelity and detection probability ($P_{YN}
\simeq 20 \%$ for the plotted case). 
We conclude that the present method is a reliable source of optical
superpositions (qubit) employing only linear components and avalanche
photodetectors. 
\begin{acknowledgments}
This work has been cosponsored by C.N.R. and N.A.T.O under the 1999 
Advanced Fellowship Program 215.31. The author thanks R. F. Antoni 
for valuable hints.
\end{acknowledgments}

\end{document}